\documentstyle[amssymb,rmp,aps]{revtex}

\begin{document}
\title{Low Temperature Relaxation of the Phase In an Inhomogeneous Bose Gas}
\author{V. S. Babichenko}
\address{R.S.C. ''Kurchatov Institute'' Kurchatov sq. 46, Moscow 123182, Russia.}
\date{\today}
\maketitle

\begin{abstract}
An effective action is obtained of a Bose gas in the bulk separated into two
regions by a strong external potential depending on the single coordinate.
The main attention is focused on the relaxation of the difference between
phases of the weakly coupling condensates of the different bulk domains
separated from each other by the external potential. The cases of low and
high temperatures are considered. 
\end{abstract}

The experimental realization of the Bose-Einstein condensation at ultralow
temperatures in atomic vapors [1] provides an example of the systems in
which the approximation of the weakly non-ideal gas is well applicable. This
fact is connected with the small density of particles in the systems
concerned. The realization of such systems gives the possibility of the
experimental investigations of the macroscopic quantum phenomenons of
different types. Recently, the manifestations of the macroscopic quantum
phase and its behavior are studied actively [2-5], [8].

The investigation of the kinetic phenomena due to the relaxation of the
order parameter is of a doubtless interest for the study of macroscopic
quantum phenomena. In the present work we study the spatial relaxation of
the phase of the order parameter in the inhomogeneous Bose gas with the weak
coupling between different spatial regions due to a barrier tunneling.

The system concerned in the present work is a Bose gas of the small density
in the bulk separated into two regions by a strong external potential $%
U\left( \overrightarrow{r}\right) $ which depends on the x coordinate and
does not depend on $\overrightarrow{r}_{\perp }=\left( y,z\right) $, i.e., $%
U\left( \overrightarrow{r}\right) =U\left( x\right) $. For simplicity, we
assume that the external potential has a rectangular shape $U\left( x\right)
=U_{0}$ for $-d<x<d$ and $U\left( x\right) =0$ beyond this region of the x
coordinate. The height of the external potential $U_{0}$ is supposed to be
the largest energy parameter in the system, in particular, $U_{0}>>\mu $
where $\mu $ is the chemical potential $\mu =n\lambda $, $\lambda $ is the
scattering amplitude of Bose particles, and $n$ is the density of the Bose
gas. Due to this assumption the interaction between particles of the Bose
gas in the region of the influence of the external potential, i.e., in the
region $-d<x<d$, can be neglected.

The left and the right domains of the bulk are supposed to have the same
temperature $T$ smaller than the Bose-condensation temperature $T_{c}$. We
assume that the densities of the right and the left domains of the bulk have
the values $n_{1}=n-\frac{1}{2}\Delta n$ and $n_{2}=n+\frac{1}{2}\Delta n$
correspondingly and the difference between densities $\Delta n$ is much
smaller than the average density n. We will consider both the case of the
nonzero $\Delta n$ and the case of $\Delta n=0$. At the same time, in these
two cases the phases of the left and the right Bose condensates are supposed
to be different at the initial time moment. This assumption means that at
the initial time moment there is a nonzero current from one side of the bulk
to another. This non-equilibrium initial state will relax to the equilibrium
state having the same densities and the same phases of the condensates for
both sides of the bulk if $\Delta n=0$ or if $\Delta n$ can change in time,
and will relax to the stationary state if the density difference $\Delta n$
keeps constant and nonzero. This relaxation process is studied in the
present work.

Note that in the case of superconductors the similar initial non-equilibrium
state results in the Josephson oscillations with a small damping [6], [7].
The essential distinction of the Bose gas consisting of neutral atoms from
superconductors in which the Cooper pairs represent the charged objects is
the presence of the gapless excitation spectrum. The presence of low energy
excitations results in the change of the character of the initial state
relaxation making the relaxation essentially faster. In the case of the
homogeneous in y-z plane difference between the phases of the condensates
the excitations radiated during the relaxation process have the
one-dimensional character. The one-dimensional character of the radiated
excitations results in a divergency of the small momentum correlators of
these excitations. Hence, the consideration of the relaxation process to the
second order in the tunneling amplitude is not correct in contrast to the
case of superconductors [7] and requires more accurate analysis [9], [14],
[15].

\section{The effective action for the difference of phases.}

Below the approach to the problem of the phase difference relaxation without
using the perturbation theory in the tunneling amplitude is developed. For
this purpose the effective action for the difference between phases of the
Bose field at the right-hand and the left-hand boundaries of the potential
barrier is derived. This effective action is obtained by the integration
over the bulk components of the Bose field with the fixed values of this
field at the potential barrier boundaries. The obtained effective action
contains the relaxation part which is proportional to the first power of the
frequency of the phase field. For the small frequencies the relaxation part
is found to be much larger than the usual kinetic part of the phase dynamics
which is proportional to the second power of the frequency. In order to
describe the phase difference relaxation process the consideration is
developed in the framework of the Schwinger-Keldysh technique [10], [11]
which is very convenient for the description of quantum kinetic processes.

The generating functional of the system can be written in the form $Z=\int
D\psi D\overline{\psi }e\exp \left\{ i\left( S_{U}+S_{j}\right) \right\} $.
The action $S_{U}$ of the inhomogeneous non-equilibrium Bose gas is given by

\[
S_{U}=\oint dt\int d^{3}r\left\{ \overline{\psi }\left[ i\partial _{t}+\mu
\left( x\right) -U\left( x\right) \right] \psi -\frac{1}{2}\left( 
\overrightarrow{\nabla }\overline{\psi }\right) \left( \overrightarrow{%
\nabla }\psi \right) -\frac{\lambda }{2}\left( \overline{\psi }\psi \right)
^{2}\right\} 
\]
and the term $S_{j}$ reads as $S_{j}=\oint dt\int d^{3}r\left( \overline{%
\psi }j+\overline{j}\psi \right) $, where $j$ and $\overline{j}$ are the
infinitely small sources. The integration over the time t in the action is
realized over the time reversing contour and is denoted as $\oint dt$. The
Planck constant $\hbar $ and mass m of a Bose particle are set equal to
unity $\hbar =m=1$. The chemical potential $\mu \left( x\right) $ is
supposed to have the constant value $\mu _{1}=\mu -\frac{1}{2}\Delta \mu $
for the coordinate x%
\mbox{$>$}%
d and $\mu _{2}=\mu +\frac{1}{2}\Delta \mu $ for x%
\mbox{$<$}%
-d where the chemical potential $\mu $ obeys the equality $\mu =\lambda n$
and the chemical potential\ difference obeys the equality $\Delta \mu
=\lambda \Delta n$. The small value of $\Delta n$ results in the small value
of the chemical potential\ difference $\Delta \mu <<\mu $.

Due to the large magnitude of potential $U_{0}$ in the region x$\in \left[
-d,d\right] $ the interaction between Bose particles can be neglected and
the integral for Z over the fields $\psi ,\overline{\psi }$ takes the
Gaussian form in this region of the x-coordinate. In this connection the
integral for Z over the fields $\psi \left( x,\overrightarrow{r}_{\perp
};t\right) ,\overline{\psi }\left( x,\overrightarrow{r}_{\perp };t\right) $
for x$\in \left[ -d,d\right] $ with the fixed magnitudes of these fields at
the boundary of the region of the nonzero external potential can be
calculated. Moreover, the characteristic scale of the $\psi $-field
variation in the time and in the $\overrightarrow{r}_{\perp }$ space is
supposed to be much larger than $1/U_{0}$ and $1/\sqrt{U_{0}}$,
respectively. It is these fields that describe the slow relaxation process
possessing the characteristic frequency proportional to the small tunneling
coefficient. Due to this assumption the tunneling amplitude of the $\psi $
fields has the local character for the time variable t and space variables $%
\overrightarrow{r}_{\perp }$. The magnitudes of the $\psi $-fields at the
boundary of the region of the nonzero external potential are denoted as $%
\psi \left( d,\overrightarrow{r}_{\perp },t\right) =\psi _{s1}\left( 
\overrightarrow{r}_{\perp },t\right) $; $\psi \left( -d,\overrightarrow{r}%
_{\perp },t\right) =\psi _{s2}\left( \overrightarrow{r}_{\perp },t\right) $.
The calculation of the integral for Z over the $\psi $-fields in the region x%
$\in \left[ -d,d\right] $ gives the action in the form of the sum of four
terms $S=S_{vol}+S_{surf}^{\left( \rho \right) }+S_{surf}^{\left(
tunn\right) }+S_{j}$. The part of the action $S_{vol}$ is

\begin{equation}
S_{vol}=\oint dt\int\limits_{U\left( x\right) =0}dx\int d^{2}r_{\perp
}\left\{ \overline{\psi }\left( i\partial _{t}+\mu \right) \psi -\frac{1}{2}%
\left( \overrightarrow{\nabla }\overline{\psi }\right) \left( 
\overrightarrow{\nabla }\psi \right) -\frac{\lambda }{2}\left( \overline{%
\psi }\psi \right) ^{2}\right\}  \eqnum{1}
\end{equation}
where the region of the integration over the x-coordinate is the sum of the
regions $\left( -L,-d\right) \cup \left( d,L\right) $. The parts of the
action $S_{surf}^{\left( \rho \right) }$ and $S_{surf}^{\left( tunn\right) }$
can be represented in the form [17] $\ $

\begin{equation}
S_{surf}^{\left( \rho \right) }=-\frac{\varkappa }{2}\coth \left( 2\varkappa
d\right) \oint dt\int d^{2}r_{\perp }\left\{ \mid \psi _{s1}\left( 
\overrightarrow{r}_{\perp },t\right) \mid ^{2}+\mid \psi _{s2}\left( 
\overrightarrow{r}_{\perp },t\right) \mid ^{2}\right\}  \eqnum{2}
\end{equation}

\begin{equation}
S_{surf}^{\left( tunn\right) }=\frac{\varkappa }{2\sinh \left( 2\varkappa
d\right) }\oint dt\int d^{2}r_{\perp }\left\{ \overline{\psi }_{s1}\left( 
\overrightarrow{r}_{\perp },t\right) \psi _{s2}\left( \overrightarrow{r}%
_{\perp },t\right) e^{i\Delta \mu t}+\overline{\psi }_{s2}\left( 
\overrightarrow{r}_{\perp },t\right) \psi _{s1}\left( \overrightarrow{r}%
_{\perp },t\right) e^{-i\Delta \mu t}\right\}   \eqnum{3}
\end{equation}
where the magnitude $\varkappa $ is $\varkappa =\sqrt{2U_{0}}$, the index 1
denotes the right-hand side of the bulk and the index 2 denotes the
left-hand side of the bulk. The term $S_{j}$ can be written as $S_{j}=\oint
dt\sum\limits_{\alpha =1,2}\left( \overline{\psi }_{s\alpha }j+\overline{j}%
\psi _{s\alpha }\right) $ where $j$ and $\overline{j}$ are the infinitely
small sources.\ Later on we assume that the width d\ of the potential
barrier is sufficiency large so that the inequality $\varkappa d>>1$ takes
place and, thus, $\sinh \left( 2\varkappa d\right) \thickapprox \cosh \left(
2\varkappa d\right) \thickapprox \frac{1}{2}e^{2\varkappa d}>>1$. The term $%
S_{surf}^{\left( tunn\right) }$ describes the tunneling between the
right-hand and left-hand sides of the bulk. From the form of this tunneling
part of the action it can easily be seen that the tunneling amplitude for
the condensate particles and for the low energy non-condensate particles has
the same nonzero magnitude in contrast with the work [8].

Our goal is the calculation of the functional integral for Z over the $\psi $%
-fields in the bulk, i.e., in the region x$\in \left( -L,-d\right) \cup
\left( d,L\right) $ with the fixed values at the potential barrier boundary,
and, thus, obtain the effective action for the fields at the boundary $\psi
_{s1}$, $\psi _{s2}$.

Below the onedimensional character of the system and its excitations is
supposed. In this supposition the considering fields depend on the x
coordinate only. Later on, it is convenient to introduce the dimensionless
space and time coordinates via the following change of these variables $%
x\rightarrow \xi x;$ \ $t\rightarrow t/\mu $. Thus, the system of units
which we use later measures the quantities of the dimension of length in
units of $\xi $ and the quantities of the dimension of energy in units of $%
\mu $. Moreover, we translate the x coordinate for the right-hand side of
the bulk on the value -d and for the left-hand side on +d, so as the
right-hand and left-hand boundaries of the potential barrier take the zero
coordinates x=0.

It is convenient to represent the fields $\psi $ in the modulus-phase form $%
\psi =Re^{i\varphi }$; \ $\overline{\psi }=Re^{-i\varphi }$. Below the
modulus $R$ of the field $\psi $ is represented as the sum of the
saddle-point configuration $F\left( x\right) $ and the fluctuations of the
modulus field $\rho $ which characterize the density fluctuations

\begin{equation}
R\left( x,t\right) =F\left( x\right) +\rho \left( x,t\right)  \eqnum{4}
\end{equation}
The saddle-point configuration $F\left( x\right) $ obeys the equation

\begin{equation}
\left( 1+\frac{1}{2}\nabla _{x}^{2}-F^{2}\left( x\right) \right) F\left(
x\right) =\varkappa F\left( x\right) \delta \left( x\right)  \eqnum{5}
\end{equation}
and has the form $F\left( x\right) =\tanh \left( \mid x\mid +X_{0}\right) $,
and the constant $X_{0}$ is the small value $X_{0}=1/\varkappa $. The
fluctuations $\rho $ can be represented as the sum of two summands

\begin{equation}
\rho \left( x,t\right) =\delta \rho \left( x,t\right) +\frac{\nabla
_{x}F\left( x\right) }{\nabla _{x}F\left( 0\right) }\rho _{s}\left( t\right)
\eqnum{6}
\end{equation}
where the fluctuations $\delta \rho \left( x,t\right) $ obey the zero
boundary conditions $\delta \rho \left( 0,t\right) =0$. Note, that the
second summand of (6) is the zero mode of the Hamiltonian $\widehat{H}_{1}=-%
\frac{1}{2}\nabla _{x}^{2}-1+3F^{2}$ which describe the dynamics of the
fluctuations $\rho \left( x,t\right) $. To describe the slow process of the
tunneling through the potential barrier we are interesting for the fields $%
\delta \rho $ and $\varphi $ which are slow varying values at the space
scale $\xi $ and the time scale $1/\mu $. For the derivation of the slow
field effective action we substitute the field $\psi $ into the action (1-3)
in the modulus-phase representation where the field $R\left( x,t\right) $
has the form (4) and where $\rho $ is determined by (6). At the same time,
in certain terms of the action due to the slowness of $\delta \rho $ and $%
\varphi $ the function like $F^{2}\left( x\right) $ can be replaced by the
function $1-\delta \left( x\right) $, where $\delta \left( x\right) $ is the 
$\delta $-function. \ Integrating over the fields $\delta \rho \left(
x,t\right) $ and $\rho _{s}\left( t\right) $ in the generation functional we
obtain the effective action for the slow phase field

\[
S^{\left( slow\right) }\left[ \varphi \right] =K\oint dt\int\limits_{U\left(
x\right) =0}dx\left\{ \frac{1}{2}\left( \partial _{t}\varphi \right) ^{2}-%
\frac{1}{2}\left( \nabla _{x}\varphi \right) ^{2}+\left[ \frac{1}{2}\left(
\partial _{t}\varphi _{s}\right) ^{2}+\gamma \cos \left( \Delta \varphi
_{s}\left( t\right) +\Delta \mu t\right) \right] \delta \left( x\right)
\right\} 
\]

where $\gamma =\frac{2}{\varkappa }\exp \left( -2\varkappa d\right) $, the
value $\Delta \varphi _{s}\left( t\right) $ is the difference of phases at
the right-hand and the left-hand boundaries of the potential barrier $\Delta
\varphi _{s}\left( t\right) =\varphi _{s2}\left( t\right) -\varphi
_{s1}\left( t\right) $, the constant $K$ is $K=n\xi S_{\perp }>>1$ and $%
\delta \left( x\right) $ is the Dirac $\delta $-function. The value $%
S_{\perp }$ is the square of the bulk section which is supposed to obey the
inequality $S_{\perp }\lesssim \xi ^{2}$ to ensure the 1D character of the
system. The dimensionless constant $K=n\xi S_{\perp }$ has the large value $%
K>>1$ due to the large value of the parameter $n\xi ^{3}>>1$. The last
inequality is ensured by the application of the gas approximation.

The last step to the derivation of the effective action for the phase
difference $\Delta \varphi _{s}\left( t\right) $ is the integration in Z \
over the bulk components of the phase $\varphi \left( x,t\right) $ with the
fixing boundary values $\varphi _{s}\left( t\right) $. For this purpose the
field $\varphi \left( x,t\right) $ is represented in the form of the Fourier
expansion $\varphi \left( x,t\right) =\sum\limits_{k}\varphi _{k}\left(
t\right) e^{ikx}$ with the Fourier components $\varphi _{k}$ obeying the
condition $\varphi _{s}\left( t\right) =\sum\limits_{k}\varphi _{k}\left(
t\right) $ where the coordinate of the right-hand boundary of the potential
barrier we choose as x=0. The integration over $\varphi _{k}$\ gives

\begin{equation}
S_{eff}\left[ \varphi _{s}\right] =\oint dtdt^{\prime }\left\{ \frac{1}{2}%
\left( \partial _{t}\varphi _{s}\right) ^{2}\delta \left( t-t^{\prime
}\right) +\frac{1}{2}\varphi _{s}\left( t\right) \left[ <\widehat{D}>\right]
_{t,t^{\prime }}^{-1}\varphi _{s}\left( t^{\prime }\right) +\gamma \cos
\left( \Delta \varphi _{s}\left( t\right) -\Delta \mu t\right) \delta \left(
t-t^{\prime }\right) \right\}  \eqnum{8}
\end{equation}
where we denote $<\widehat{D}>=\int \frac{dk}{2\pi }\widehat{D}_{k}$. In the
frequency-momentum representation in the ''triangular'' form [10], [11] at
the temperature T the phonon propagator $\widehat{D}\left( \omega ,k\right) $
can be written

\begin{equation}
\widehat{D}\left( \omega ,k\right) =\left( 
\begin{array}{cc}
0 & D^{A}\left( \omega ,k\right) \\ 
D^{R}\left( \omega ,k\right) & D^{K}\left( \omega ,k\right)
\end{array}
\right)  \eqnum{9}
\end{equation}
where $D^{R,A}\left( \omega ,k\right) =\left[ \left( \omega \pm i\delta
\right) ^{2}-k^{2}\right] ^{-1}$ and $D^{K}\left( \omega ,k\right) =\coth
\left( \frac{\omega }{2T}\right) \left( D^{R}-D^{A}\right) $. The simple
calculations of $<\widehat{D}>_{\omega }$ give

\begin{equation}
<\widehat{D}>_{\omega }=\frac{1}{2}\left( 
\begin{array}{cc}
0 & <D^{A}>_{\omega } \\ 
<D^{R}>_{\omega } & <D^{K}>_{\omega }
\end{array}
\right) \text{\ }  \eqnum{10}
\end{equation}
where $<D^{R}>_{\omega }=\frac{1}{i\left( \omega +i\delta \right) }$; $%
<D^{A}>_{\omega }=\frac{1}{-i\left( \omega -i\delta \right) }$; $\
<D^{K}>_{\omega }=2\coth \left( \frac{\omega }{2T}\right) \frac{1}{i\omega }$%
.

\section{The effective action for small frequencies and the Fokker-Planck
equation.}

Introducing the half-sum and the difference of the fields instead of the
fields $\varphi _{s1}$ and $\varphi _{s2}$ we obtain the effective action
for the difference of phases at the boundary of the potential barrier. In
the ''triangle'' representation it can be written as

\begin{equation}
S\left[ \varphi \right] =2K\int dt\left\{ \varphi \left[ \frac{1}{2}\widehat{%
\omega }^{2}+i\widehat{\omega }\right] \Phi +\Phi \left[ \frac{1}{2}\widehat{%
\omega }^{2}-i\widehat{\omega }\right] \varphi +i\varphi \left[ 2\widehat{%
\omega }\coth \left( \frac{\widehat{\omega }}{2T}\right) \right] \varphi -V%
\left[ \Phi ,\varphi \right] \right\}  \eqnum{11}
\end{equation}
where $\widehat{\omega }=i\partial _{t}$, the function $V\left[ \Phi
,\varphi \right] $ is $V\left[ \Phi ,\varphi \right] =\gamma \sin \left(
\Phi +\Delta \mu t\right) \sin \left( \varphi \right) $. In this
representation the fields $\Phi $ and $\varphi $ are expressed via the
fields belonging to the upper and the lower time contour branches as $\Phi =%
\frac{1}{2}\left( \varphi _{+}+\varphi _{-}\right) $ and $\varphi =\frac{1}{2%
}\left( \varphi _{+}-\varphi _{-}\right) $. Note that as usually in the
''triangle'' representation of the Keldysh-Schwinger technique the field $%
\Phi $ describes the kinetics of the phase difference and the field $\varphi 
$ describes the quantum noise of this value.

The fields $\Phi $ and $\varphi $ can be fragmented into fast and slow
components. The slow fields are the fields having the frequencies $\omega $
obeying the inequality $\omega <\omega _{0}$ and the fast fields are the
fields with the frequencies $\omega >\omega _{0}$ where $\gamma <<\omega
_{0}<<1$. The part of the action for the fast fields is obtained by the
expansion of the potential $V$ \ up to the second order of the fast
components $\delta \Phi $, $\delta \varphi $ and has the form

\begin{equation}
S^{\left( fast\right) }\left[ \delta \varphi ,\delta \Phi \right] =2K\int
dt\left\{ 
\begin{array}{c}
\delta \varphi \left[ \frac{1}{2}\widehat{\omega }^{2}+i\widehat{\omega }%
\right] \delta \Phi +\delta \Phi \left[ \frac{1}{2}\widehat{\omega }^{2}-i%
\widehat{\omega }\right] \delta \varphi +i\delta \varphi \left[ 2\widehat{%
\omega }\coth \left( \frac{\widehat{\omega }}{2T}\right) \right] \delta
\varphi - \\ 
-V_{\varphi \Phi }^{\prime \prime }\delta \varphi \delta \Phi -\frac{1}{2}%
V_{\Phi \Phi }^{\prime \prime }\left( \delta \Phi \delta \Phi +\delta
\varphi \delta \varphi \right)
\end{array}
\right\}  \eqnum{12}
\end{equation}

\[
V_{\varphi \Phi }^{\prime \prime }=\gamma \cos \left( \Phi +\Delta \mu
t\right) \cos \left( \varphi \right) \text{; \ \ \ }V_{\Phi \Phi }^{\prime
\prime }=V_{\varphi \varphi }^{\prime \prime }=-\gamma \sin \left( \Phi
+\Delta \mu t\right) \sin \left( \varphi \right) \text{\ } 
\]
where $\delta \varphi ,\delta \Phi $ are the fast components of fields and $%
\varphi ,\Phi $ are the slow components. The part of the action for the slow
fields can be written as

\begin{equation}
S^{\left( slow\right) }\left[ \varphi \right] =2K\int dt\left\{ -2\varphi
\partial _{t}\Phi -V\left[ \Phi ,\varphi \right] +i\varphi \left[ 2\widehat{%
\omega }\coth \left( \frac{\widehat{\omega }}{2T}\right) \right] \varphi
\right\}  \eqnum{13}
\end{equation}
Due to the large value of the constant $K>>1$, so as the value $\frac{1}{K}%
\ln \frac{1}{\gamma }$ obeys the inequality $\frac{1}{K}\ln \frac{1}{\gamma }%
<<1$, the action $S^{\left( fast\right) }\left[ \delta \varphi ,\delta \Phi %
\right] $ can be approximated as

\begin{equation}
S^{\left( fast\right) }\left[ \delta \varphi ,\delta \Phi \right] =2K\int
dt\left\{ 
\begin{array}{c}
\delta \varphi \left[ \frac{1}{2}\widehat{\omega }^{2}+i\widehat{\omega }%
\right] \delta \Phi +\delta \Phi \left[ \frac{1}{2}\widehat{\omega }^{2}-i%
\widehat{\omega }\right] \delta \varphi +i\delta \varphi \left[ 2\widehat{%
\omega }\coth \left( \frac{\widehat{\omega }}{2T}\right) \right] \delta
\varphi - \\ 
-V_{\varphi \Phi }^{\prime \prime }<\delta \varphi \delta \Phi >_{0}-\frac{1%
}{2}V_{\Phi \Phi }^{\prime \prime }<\delta \Phi \delta \Phi >_{0}-\frac{1}{2}%
V_{\varphi \varphi }^{\prime \prime }<\delta \varphi \delta \varphi >_{0}
\end{array}
\right\}  \eqnum{14}
\end{equation}
where

\begin{eqnarray}
&<&\delta \varphi _{t}\delta \varphi _{t}>_{0}=0\text{; }<\delta \Phi
_{t}\delta \Phi _{t}>_{0}=\frac{i}{2K}\int\limits_{\mid \omega \mid >\omega
_{0}}\frac{d\omega }{2\pi }D^{\left( K\right) }\left( \omega \right) =\frac{1%
}{4\pi K}\ln \left( \frac{1}{\omega _{0}}\right) \text{; }  \eqnum{15} \\
&<&\delta \varphi _{t}\delta \Phi _{t}>_{0}=i\int\limits_{\mid \omega \mid
>\omega _{0}}\frac{d\omega }{2\pi }D_{0}^{\left( R\right) }\left( \omega
\right) =\frac{i}{8K}  \nonumber
\end{eqnarray}
The correlator $\widehat{D}_{0}$ is defined by the equality $\widehat{D}%
_{0}=\left( \frac{\omega ^{2}}{2}\widehat{\sigma }_{x}+\frac{1}{2}<\widehat{D%
}>^{-1}\right) ^{-1}$, where $\widehat{\sigma }_{x}$ is the $\widehat{\sigma 
}_{x}$-\ Pauli matrix. Thus, taking into account the renormalizations
resulting from the fast fluctuations the effective action for the slow
fields can be written as

\begin{equation}
S^{\left( slow\right) }\left[ \varphi ,\Phi \right] =2K\int dt\left\{
-2\varphi \left( \partial _{t}\Phi \right) -V\left[ \Phi ,\varphi \right]
+i\varphi \left[ 2\widehat{\omega }\coth \left( \frac{\widehat{\omega }}{2T}%
\right) \right] \varphi -\frac{i}{8K}V_{\varphi \Phi }^{\prime \prime
}\right\}  \eqnum{16}
\end{equation}
In this expression the renormalizations proportional to the correlators $%
<\delta \Phi _{t}\delta \Phi _{t}>_{0}$ give the small renormalization of
the constant $\gamma $ in the potential $V$. They can be neglected due to
the smallness of the value $\frac{1}{K}\ln \left( \frac{1}{\omega _{0}}%
\right) <<1$. On the other side the renormalizations proportional to the
correlators $<\delta \varphi _{t}\delta \Phi _{t}>_{0}$ are taken into
account because they can become comparable with the relaxation term $\varphi %
\left[ 2\omega \coth \left( \frac{\omega }{2T}\right) \right] \varphi $ in
the case of the low temperatures and small frequencies $T\thicksim \frac{1}{K%
}$, $\omega \thicksim \frac{1}{K}$.

It is convenient to shift the field $\Phi \rightarrow \Phi -\Delta \mu t$.
As a result of this shift the action (16) takes the form

\[
S^{\left( slow\right) }\left[ \varphi ,\Phi \right] =2K\int dt\left\{
-2\varphi \left( \partial _{t}\Phi \right) -V^{\left( 0\right) }\left[ \Phi
,\varphi \right] +i\varphi \left[ 2\widehat{\omega }\coth \left( \frac{%
\widehat{\omega }}{2T}\right) \right] \varphi -\frac{i}{8K}V_{\varphi \Phi
}^{\left( 0\right) \prime \prime }\right\} 
\]
where $V^{\left( 0\right) }\left[ \Phi ,\varphi \right] =\gamma \sin \Phi
\sin \left( \varphi \right) -v_{0}\varphi $, at that $v_{0}=2\Delta \mu $.
Taking into account the smallness of the quantum noise $\varphi $ due to the
large value of the constant $K>>1$ the action can be expanded in the powers
of the field $\varphi $. This expansion gives

\begin{equation}
S_{eff}^{\left( slow\right) }\left[ \varphi ,\Phi \right] =2K\int dt\left\{
-2\varphi \left( \partial _{t}\Phi \right) +i\varphi \left[ 2\widehat{\omega 
}\coth \left( \frac{\widehat{\omega }}{2T}\right) +\frac{1}{16K}\gamma \cos
\Phi \right] \varphi -A\left( \Phi \right) \varphi -i\frac{1}{8K}A^{\prime
}\left( \Phi \right) \right\}  \eqnum{17}
\end{equation}
where

\begin{equation}
A\left( \Phi \right) =\gamma \sin \Phi -v_{0}\text{; \ \ \ }A^{\prime
}\left( \Phi \right) =\gamma \cos \Phi \text{\ }  \eqnum{18}
\end{equation}
The first correction to the renormalized amplitude of the potential $V\left[
\Phi ,\varphi \right] $, i.e., to the constant $\gamma $ with respect to the
value $1/K$ has the form $\gamma ^{\ast }=\gamma \left( 1-\frac{1}{4\pi K}%
\ln \left( \frac{K}{\gamma }\right) \right) $. Integrating over $\varphi $
in the generation functional Z we obtain

\[
S_{eff}^{\left( slow\right) }\left[ \Phi \right] =-\frac{1}{2}Sp\ln \left( 
\widehat{\Gamma }\right) +2K\int dtdt^{\prime }\left\{ \frac{i}{4}\left[
-2\left( \partial _{t}\Phi \right) -A_{t}\right] \widehat{\Gamma }%
_{t,t^{\prime }}\left[ -2\left( \partial _{t^{\prime }}\Phi \right)
-A_{t^{\prime }}\right] -i\frac{1}{8K}A_{t}^{\prime }\delta \left(
t-t^{\prime }\right) \right\} 
\]
where the correlator $\widehat{\Gamma }_{t,t^{\prime }}$ is

\begin{equation}
\widehat{\Gamma }=\left[ 2\widehat{\omega }\coth \left( \frac{\widehat{%
\omega }}{2T}\right) +\frac{1}{16K}\gamma ^{\ast }\cos \Phi \right]
^{-1}\approx \left[ 2\widehat{\omega }\coth \left( \frac{\widehat{\omega }}{%
2T}\right) +\frac{1}{16K}\gamma ^{\ast }\right] ^{-1}  \eqnum{19}
\end{equation}
In the approximation of the local in time action we have

\begin{equation}
S_{eff}^{\left( slow\right) }\left[ \Phi \right] =i2K\int dt\left\{ \frac{1}{%
4}\left[ 2\left( \partial _{t}\Phi _{t}\right) +A\left( \Phi _{t}\right) %
\right] ^{2}\widehat{\Gamma }_{0}-\frac{1}{8K}A^{\prime }\left( \Phi
_{t}\right) \right\}  \eqnum{20}
\end{equation}

\begin{equation}
\widehat{\Gamma }_{0}=\frac{1}{4T+\frac{1}{16K}\gamma ^{\ast }}=\frac{1}{%
4T^{\ast }}  \eqnum{21}
\end{equation}

Finally, the action can be represented in the form

\begin{equation}
S_{eff}^{\left( slow\right) }\left[ \Phi \right] =i2K\int dt\left\{ \frac{M}{%
2}\left( \partial _{t}\Phi \right) ^{2}+\frac{M}{8}A^{2}-\frac{1}{8K}%
A^{\prime }\right\}  \eqnum{22}
\end{equation}
where $M=2\widehat{\Gamma }_{0}=\frac{1}{2T^{\ast }}$. Note that the
generation potential Z with the action $S_{eff}^{\left( slow\right) }\left[
\Phi \right] $ (22) has the form corresponding to the quantum mechanics with
the ''imaginary time'' in which the role of the ''imaginary time'' plays the
usual real time. This action corresponds to the action which is obtained for
the classical random process describing by the Langevin equation with the
white noise [14] and having the form of the supersymmentric quantum
mechanics. The Schredinger equation and the Hamiltonian corresponding to the
action $S_{eff}^{\left( slow\right) }\left[ \Phi \right] $ are

\begin{equation}
-\frac{1}{2K}\partial _{t}\Psi =\widehat{H}_{FP}\Psi  \eqnum{23}
\end{equation}

\begin{equation}
\widehat{H}_{FP}=-\frac{1}{\left( 2K\right) ^{2}}\frac{d^{2}}{2Md^{2}\Phi }+%
\frac{M}{8}A^{2}-\frac{1}{8K}A^{\prime }  \eqnum{24}
\end{equation}
This Hamiltonian $\widehat{H}_{FP}$ can be represented in the factorized form

$\widehat{H}_{FP}=-\frac{1}{2M\left( 2K\right) ^{2}}\left( \frac{d}{d\Phi }%
-MKA\right) \left( \frac{d}{d\Phi }+MKA\right) $. The transformation $\Psi
=\exp \left\{ MKV\left( \Phi \right) \right\} P\left( \Phi \right) $, where $%
V\left( \Phi \right) =\gamma ^{\ast }\left( 1-\cos \Phi \right) -v_{0}\Phi $
and $A=\frac{d}{d\Phi }\left( V\left( \Phi \right) \right) $ transfer the
Hamiltonian (18) to the form of the Fokker-Planck operator $\widehat{L}%
_{FP}=-\frac{1}{2M\left( 2K\right) ^{2}}\frac{d}{d\Phi }\left( \frac{d}{%
d\Phi }+2MKA\right) $. The stationary solution of the corresponding
Fokker-Planck equation in the case of the zero velocity $v_{0}=0$\ has the
form $P\left( \Phi \right) =C_{0}\exp \left( -\frac{KV\left( \Phi \right) }{%
T^{\ast }}\right) $, where $V\left( \Phi \right) =\gamma ^{\ast }\left(
1-\cos \Phi \right) $ and the constant $C_{0}$ is $C_{0}=\left( \frac{%
K\gamma ^{\ast }}{2\pi T^{\ast }}\right) ^{\frac{1}{2}}$. In the case of the
nonzero velocity $v_{0}$ the stationary solution of the Fokker-Planck
equation is determined by the equation

\[
\left( \frac{d}{d\Phi }+2MKA\right) P\left( \Phi \right) =j 
\]
and can be found taking into account the condition of the periodicity of the
probability distribution $P\left( 0\right) =P\left( \Phi \right) $ and the
normalization condition $\int\limits_{0}^{2\pi }d\Phi P\left( \Phi \right)
=1 $ [13], [18]. The constant j is found from the normalization condition
for $P\left( \Phi \right) $ and for the case of the small $\Delta n$ we
obtain $j=\frac{v_{0}}{\gamma ^{\ast }}\exp \left\{ -2\left( \frac{K\gamma
^{\ast }}{T^{\ast }}\right) \right\} $. The probability of the phase slip
process is determined by the value $\left( \partial _{t}<\Phi >\right) $ $%
\sim j$. For the zero temperature and small velocities $v_{0}$, so as $%
v_{0}=\Delta \mu <<T^{\ast }$, the value $T^{\ast }$ can be considered as
the effective temperature and for the large time t, when the action can be
considered in the local in time form, determines the diffusion coefficient.
In the opposite case $v_{0}=\Delta \mu >>T^{\ast }$ the diffusion
coefficient is defined by the value $\widehat{\Gamma }_{t,t}=\ln \left( 
\frac{1}{v_{0}}\right) $\ and this value should be substituted instead of
the value $\left( \gamma ^{\ast }/T^{\ast }\right) $. This can be seen from
the expression (20) for $\widehat{\Gamma }$ with taking into account that in
the case of the large value of $\Delta \mu $ compared with $T^{\ast }$ the
frequencies $\omega $ should be cut of for the small values by the chemical
potential difference $\Delta \mu $. Note, that the result for the
probability of the phase slip $\left( 1/\tau _{ps}\right) \thicksim
v_{0}\exp \left\{ -2K\ln \left( \frac{1}{v_{0}}\right) \right\} $ coincides
with the result of the works [15-17].

The Fokker-Planck equation describing the relaxation process of the
difference of phases has been obtained both for high temperatures and for
low temperatures. For high temperatures $T\gtrsim \gamma $ the obtained
Fokker-Planck equation is equal to the classical Langevin equation with the
white noise with the correlator proportional to 1/T [18]. For the low
temperatures $T<<\gamma $ and the small density difference $\Delta n$, so as
the inequality $v_{0}<<1/K$ takes place, the relaxation kinetics of the
difference of phases is described by the Fokker-Planck equation with the
effective temperature $T^{\ast }$ and the diffusion coefficient determined
by both the usual temperature and the quantum fluctuations of the Bose gas.
For the small temperatures and the large value of the density difference $%
\Delta n$, so as $v_{0}>>\gamma /K$, the diffusion coefficient is determined
by the value $v_{0}$. In the last case the result is analogues to the result
of the works [15-17].

The author thanks S. Burmistrov, A. Kozlov and Yu. Kagan for helpful
discussions. This work was supported by the Russian Foundation for Basic
Research, by the Netherlands Organization for Scientific Research (NWO) and
by INTAS -2001-2344.

\end{document}